\documentclass[12pt]{article}
\usepackage{graphicx}
\usepackage{subfigure, epsfig, rotate, epsf,setspace, amssymb, amsxtra, placeins,slashed}
\usepackage[hmargin=2.5cm,vmargin=2.5cm]{geometry}

\newcommand{\prop}{\mathcal{G}}
\newcommand\pubnumber{DPF2013-154}
\newcommand\pubdate{\today}

\def\support{\footnote{Work supported by the US Department of Energy, the UK STFC and Japan's RIKEN institute.}}

\def\Title#1{\begin{center} {\Large #1 } \end{center}}
\def\Author#1{\begin{center}{ \sc #1} \end{center}}
\def\Address#1{\begin{center}{ \it #1} \end{center}}

\newcommand\pubblock{\rightline{\begin{tabular}{l} \pubnumber\\
         \pubdate  \end{tabular}}}
\newenvironment{Abstract}{\begin{quotation}  }{\end{quotation}}
\newenvironment{Presented}{\begin{quotation} \begin{center} 
             PRESENTED AT\end{center}\bigskip 
      \begin{center}\begin{large}}{\end{large}\end{center} \end{quotation}}

\def\beq{\begin{equation}}
\def\eeq#1{\label{#1}\end{equation}}
\def\eeqn{\end{equation}}

\def\beqa{\begin{eqnarray}}
\def\eeqa#1{\label{#1}\end{eqnarray}}
\def\eeqan{\end{eqnarray}}

\let\bar=\overbar

\def\Dslash{\not{\hbox{\kern-4pt $D$}}}
\def\dslash{\not{\hbox{\kern-2pt $\del$}}}

\def\msb{{\bar{\ssstyle M \kern -1pt S}}}

\begin{document}
\begin{titlepage}
\pubblock

\vfill
\Title{Progress Towards an {\it ab initio}, Standard Model Calculation of Direct CP-Violation in K-decays}
\vfill
\Author{ Christopher Kelly\support}
\Address{
        RIKEN-BNL Research Center,\\
        Brookhaven National Laboratory,\\
        Upton, NY 11973, USA.\\
        E-mail: \textit{ckelly@bnl.gov}
}
\vfill
\begin{Abstract}
We discuss the RBC\&UKQCD collaboration's progress towards a first-principles calculation of direct CP-violation in the Standard Model via $K\rightarrow \pi \pi$ decays. In particular we focus upon the calculation of the $I=0$ channel amplitude $A_0$, for which obtaining physical kinematics requires more sophisticated techniques than those used for the $I=2$-channel decay. We discuss our chosen techniques along with preliminary demonstrations of their application to simpler lattice quantities, and finally discuss our progress in generating the large-volume, physical-pion-mass ensembles that will be used to perform the $A_0$ calculation.
\end{Abstract}
\vfill
\begin{Presented}
DPF 2013\\
The Meeting of the American Physical Society\\
Division of Particles and Fields\\
Santa Cruz, California, August 13--17, 2013\\
\end{Presented}
\vfill
\end{titlepage}

\section{Introduction}
\vspace{-0.3cm}
Direct CP-violation in $K\rightarrow\pi\pi$ decays manifests as a difference in phase between the complex amplitudes, $A_2$ and $A_0$, of the decay in the $I=2$ ($\Delta I=3/2$) and $I=0$ ($\Delta I=1/2$) channels respectively (the $I=1$ channel being forbidden by Bose symmetry):
\vspace{-0.15cm}
\begin{equation}
\epsilon^\prime = \frac{i\omega e^{i(\delta_2-\delta_0)}}{\sqrt 2} \left(\frac{{\rm Im}A_2}{{\rm Re}A_2} - \frac{{\rm Im}A_0}{{\rm Re}A_0}\right)\,,
\vspace{-0.20cm}
\end{equation} 
where $\omega = {\rm Re}A_2/{\rm Re}A_0$ and $\delta_i$ are the scattering phase shifts of the final-state pions. 

Low energy strong interactions play an important role in the dynamics of these decays; for example they are largely responsible~\cite{Boyle:2012ys} for the factor of $\sim 20$ relative enhancement of the $I=0$ decay amplitude relative to the $I=2$ amplitude that is known as the ``$\Delta I=1/2$-rule". As a result we must use lattice QCD in order to study these processes. As the hadronic scale of $\sim 1$ GeV is much smaller than the W-boson mass, the decay can be described using the weak effective theory. The interaction takes the form of a local operator:
\vspace{-0.4cm}
\begin{equation}
H_W^{\rm eff} = \frac{G_F}{\sqrt{2}} V_{us}^* V_{ud}\sum_{i=1}^{10} \left[ z_i(\mu) + \tau y_i(\mu)\right] Q_i(\mu)\,,
\vspace{-0.3cm}
\end{equation}
where $z_i$ and $y_i$ are Wilson coefficients determined in the perturbative regime and $\tau = -V_{ts}^*V_{td}/V_{ud}V_{us}^*$ is responsible for the direct CP-violation in the decay. On the lattice we measure Euclidean Green's functions, $\langle \pi\pi| Q_i | K\rangle$, and non-perturbatively renormalize at the scale $\mu$.

The first-principles calculation of $\epsilon'$ has long been a goal of the lattice community, but it is only recently that the techniques and raw computing power have become available to perform a realistic calculation. The difficulties are two-fold: firstly, performing the calculation requires both large physical volumes and light quark masses, which in turn requires large and powerful supercomputers; and secondly it requires the development of strategies for calculating diagrams with vacuum intermediate states (in the $I=0$ channel), and for obtaining the physical kinematics in the decay.

Using large-volume but relatively coarse ensembles of domain wall fermions with the Iwasaki+DSDR gauge action and near-physical pion masses, the RBC and UKQCD collaboration have performed the first realistic {\it ab initio} calculation of the decay amplitude in the $I=2$ channel~\cite{Blum:2011ng,Blum:2012uk}, and a calculation with finer lattices and a full continuum extrapolation is underway that will substantially reduce the discretization systematic, which was the largest contribution to the error on the earlier calculation. Unfortunately the techniques used to obtain physical kinematics in this channel (discussed further below) are not applicable to the $I=0$ case, and an alternative strategy must be found. For this we have chosen G-parity boundary conditions, the discussion of which will be the focus of these proceedings.

\vspace{-0.5cm}
\section{Obtaining physical kinematics}
\vspace{-0.3cm}
%Lattice calculations are necessarily performed in a finite volume. For multi-particle states, the component particles are typically confined in close proximity, and their continuous interaction causes changes in the state's energy and normalization that are functions of the physical scattering length. We must therefore calculate an additional quantity, known as the Lellouch-Luscher factor~\cite{Lellouch:2000pv}, which relates the finite-volume amplitudes to their physical, infinite-volume values. The determination of this quantity demands that the interaction is on-shell. 

We are interested in measuring the on-shell, physical decay. As the kaon mass is $500$ MeV and the pion mass is $135$ MeV, this requires the final state pions to each have non-zero momentum. This is an excited state of the $\pi\pi$-system, the ground state of which comprises stationary pions. Lattice calculations are performed by measuring Euclidean Green's functions, the time-dependence of which can in general be described as an infinite sum of terms decaying exponentially in the energy of 
each allowed state. The cleanest signal is usually the ground-state contribution, which can be picked out in the large-time limit. In principle it is possible to extract excited state contributions providing one has sufficient statistical precision to resolve them over the much larger ground-state contribution. However for this calculation it is unlikely that a precise-enough measurement could be performed using this strategy, particularly in the $I=0$ case where the presence of vacuum diagrams is expected to lead to considerably noisier measurements. 

For the $\Delta I=3/2$ measurement it is possible to modify the quark boundary conditions in order to induce momentum on the final state pions: With antiperiodic valence boundary conditions (BC), the finite-volume discretization of the lattice momentum changes from integer multiples of $2\pi/L$ with periodic BC to odd-integer multiples of $\pi/L$, where $L$ is the lattice spatial side length. Imposing these conditions on the down-quark propagator while retaining periodic BC for the up quark results in a charged pion state that is antiperiodic: $\pi^+(x+L) = \bar d(x+L)u(x+L)= -\bar d(x)u(x)$. However, the neutral pion, which is needed for the physical decay $K^+\rightarrow \pi^+\pi^0$, remains periodic as $\bar d(x+L)d(x+L)= \bar d(x)d(x)$. This can be avoided using the Wigner-Eckart theorem to relate the physical decay to an unphysical one containing only charged pions:
\vspace{-0.6cm}
\begin{equation}
\langle (\pi^+\pi^0)_{I=2}| Q^{\Delta I_z = 1/2} | K^+\rangle = \frac{\sqrt{3}}{2}\langle (\pi^+\pi^+)_{I=2} | Q^{\Delta I_z = 3/2} | K^+\rangle\,.
\vspace{-0.4cm}
\end{equation}
This trick circumvents another issue; that imposing different boundary conditions on the up and down quarks manifestly breaks the isospin symmetry, allowing mixing between states of different isospin. This is prevented here by charge conservation, as the final state is the only charge-2 state that can be formed with the remaining quantum numbers. Using this technique, the RBC and UKQCD collaboration were able to obtain $A_2$~\cite{Blum:2011ng,Blum:2012uk}.

Unfortunately the strategy described above cannot be employed for the calculation of $A_0$: This calculation requires the measurement of $K^0\rightarrow \pi^+\pi^-$ and also $K^0\rightarrow \pi^0\pi^0$, where for $I=0$ there is no convenient Wigner-Eckart relation to remove the neutral pions from the latter. There is also no means of avoiding the isospin-breaking induced by imposing different BC on the down and up quarks. G-parity boundary conditions (GPBC)~\cite{Wiese:1991ku, Kim:2003xt, Kim:2009fe} offer a means to circumvent these issues.
\vspace{-0.5cm}
\section{G-parity Boundary Conditions}
\vspace{-0.3cm}
G-parity is a combination of charge conjugation and an isospin rotation by $\pi$ radians about the y-axis: $\hat G = \hat C e^{i\pi \hat I_y}$, where the hat-symbol is used to denote operators. The charged {\it and} neutral pions are all eigenstates of this operation with eigenvalue $-1$, hence applying the operation at a spatial boundary causes the pion states to become antiperiodic in that direction, removing the zero-momentum ground state. 

At the quark level,
\vspace{-0.3cm}
\begin{equation}
\hat G\left(\begin{array}{c}u\\d\end{array}\right) = \left(\begin{array}{c}-C\bar d^T\\C\bar u^T\end{array}\right)\,,
\vspace{-0.2cm}
\end{equation}
where $C=\gamma^2\gamma^4$ in our conventions. Due to the mixing between the quark flavours at the lattice boundary, extensive modifications to our parallel code libraries were required to perform an efficient calculation. 

\begin{figure}[t]
\centering
\includegraphics*[width=0.45\textwidth]{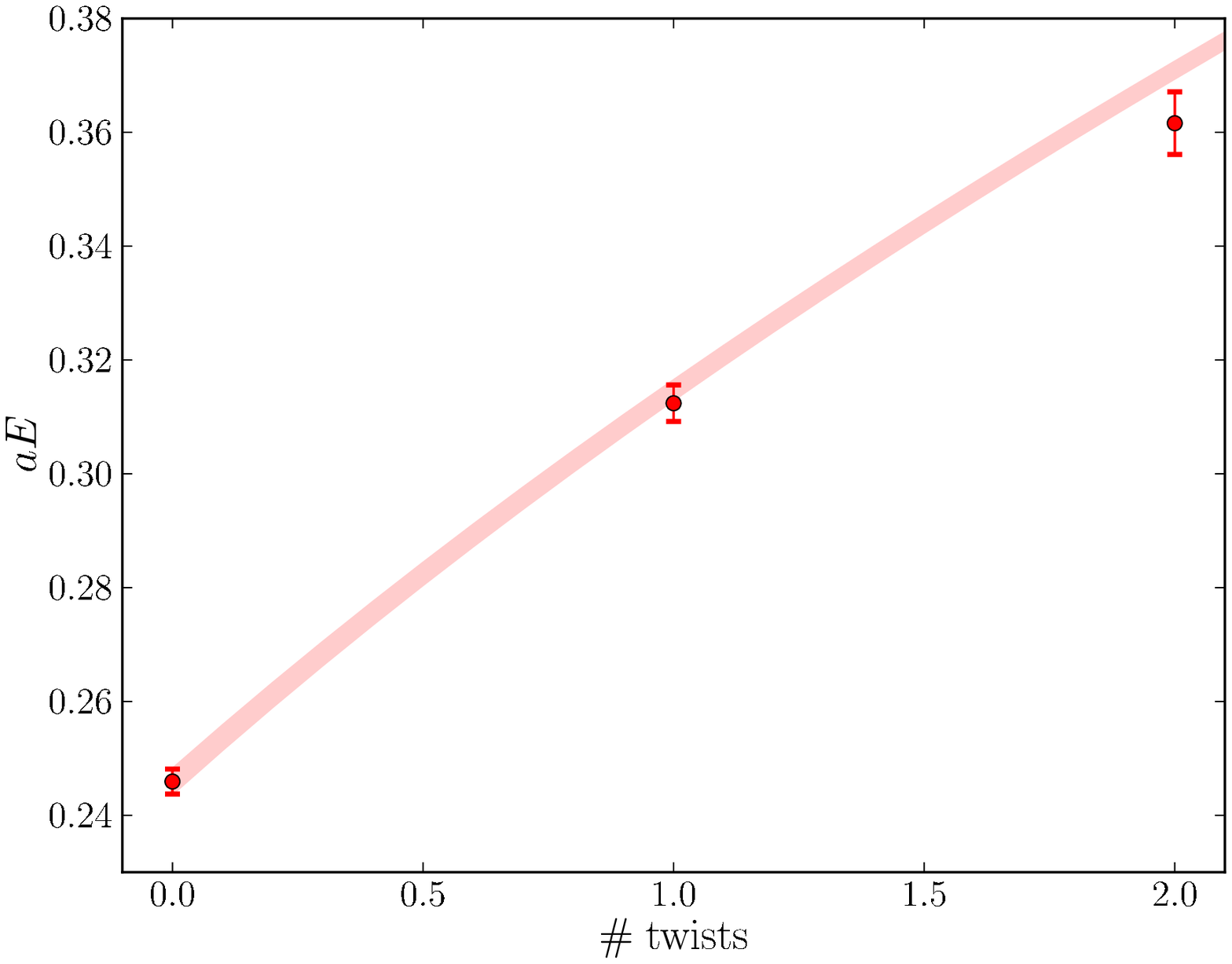}
\includegraphics*[width=0.45\textwidth]{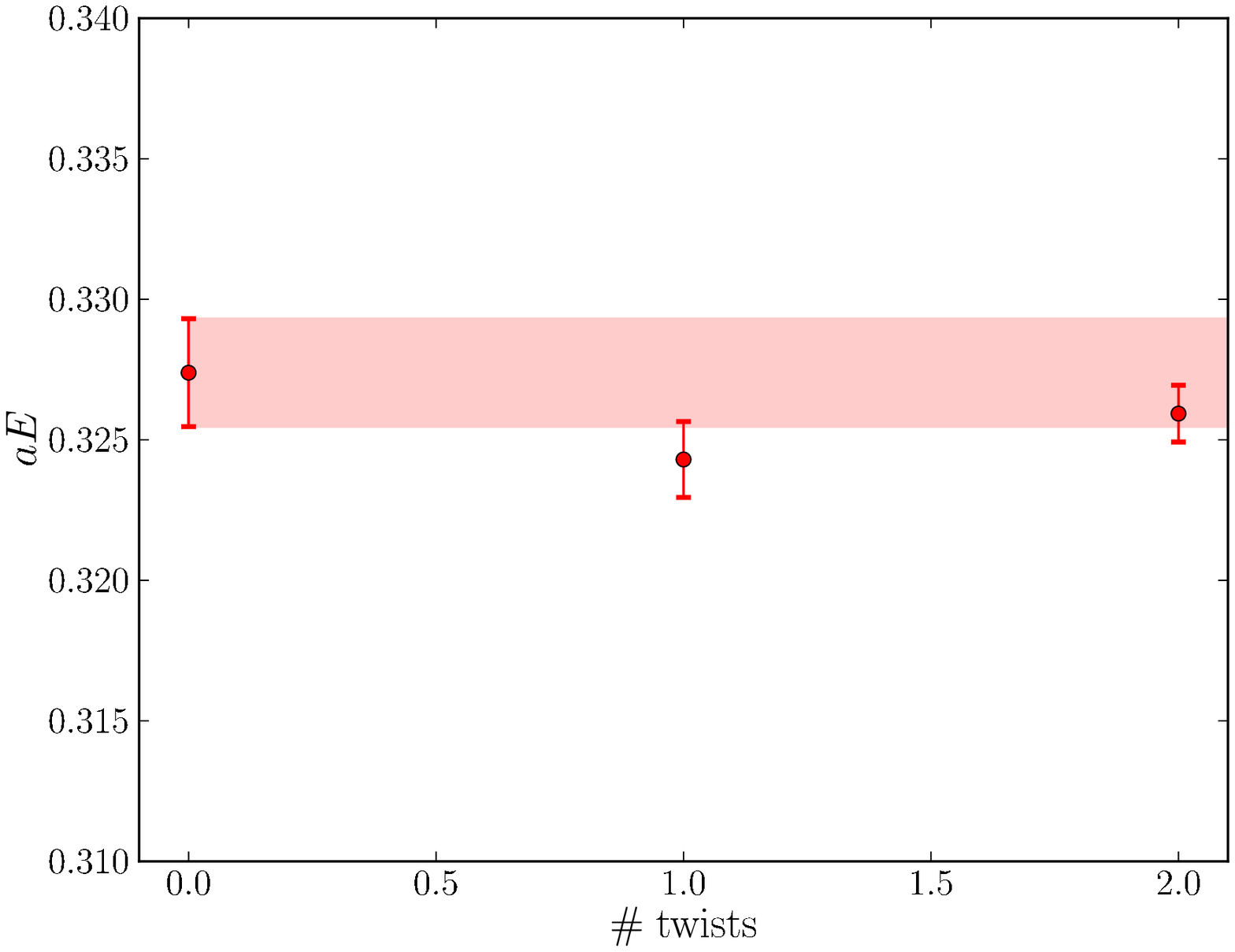}
\includegraphics*[height=0.45\textwidth,angle=-90,trim=0 0 0 0,clip=true]{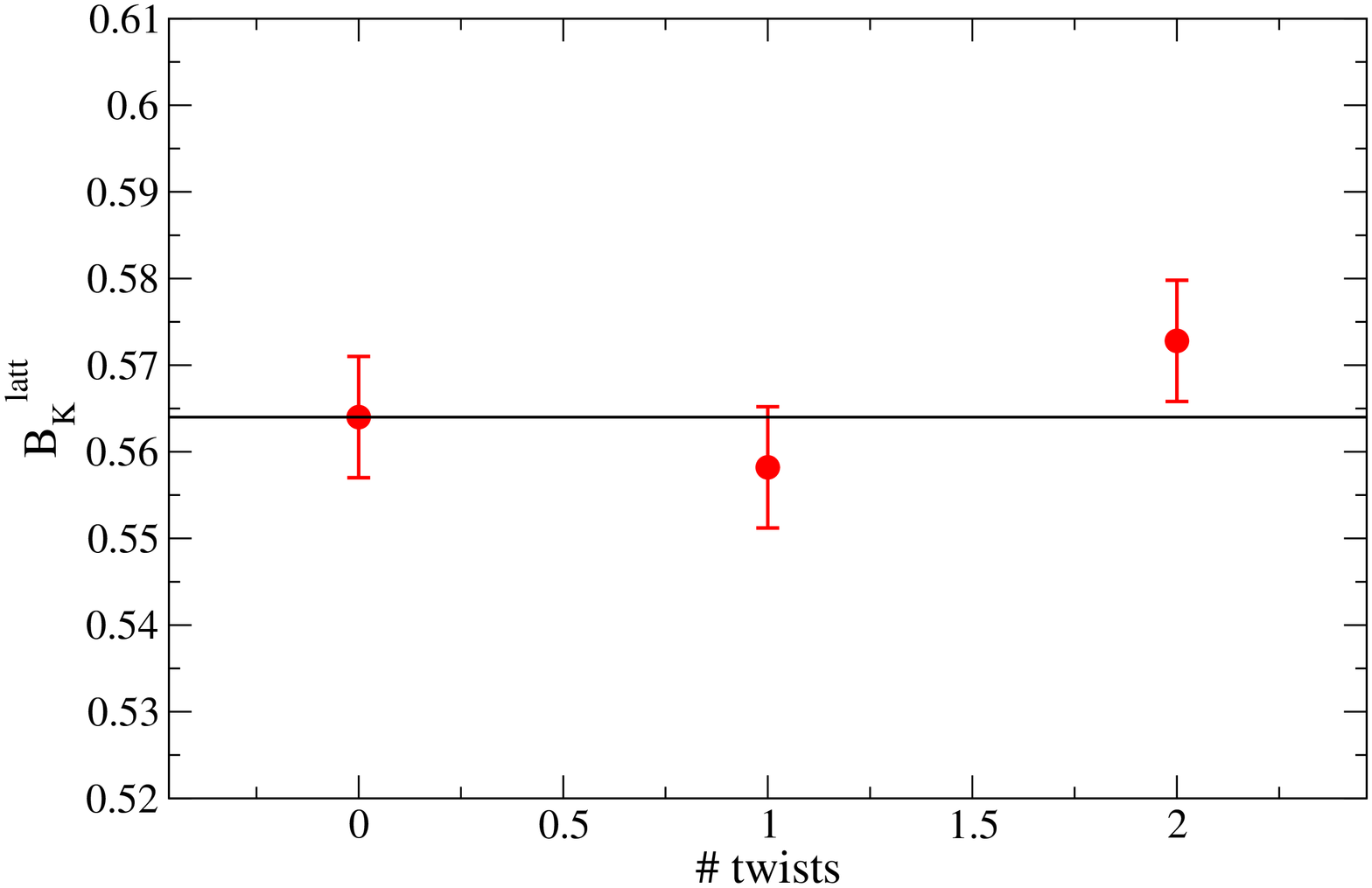}
\caption{\label{fig-dispersion}Top: the measured pion and kaon energies respectively as a function of the number of G-parity directions (twists), overlaid by the expected continuum dispersion relations. Bottom: $B_K$ as a function of the number of G-parity directions.}
\end{figure}

Additional complications arise from the fact that the Dirac operator for the fields across the boundary involves complex conjugated gauge links -- the ${\rm SU}(3)$ matrices describing the gluon and sea-quark dynamics -- which necessitates the generation of new ensembles of gauge configurations obeying complex conjugate BC, an unusual and expensive requirement for a lattice calculation. (Note that using antiperiodic BCs in this calculation would also require new ensembles to be generated due to the presence of vacuum diagrams.)

In order to describe states involving strange quarks, in particular the stationary neutral kaons required for the $A_0$ calculation, with the quarks interacting with gauge fields that obey complex-conjugate BC, we place the strange quark in an isospin doublet with a fictional degenerate partner, referred to as $s'$, and impose GPBC on this pair. We can then form a state comprising the usual kaon and a fictional particle involving the partner quarks: $\tilde K = \frac{1}{\sqrt{2}}( \bar s d + \bar u s' )$. This is an eigenstate of G-parity with eigenvalue $+1$, and thus obeys periodic BC and has a stationary ground state. If we restrict ourselves to operators involving only the physical strange quark, then the fictional partner to the kaon can only contribute by propagating through the boundary, an effect which is suppressed exponentially in the lattice size and the kaon mass, and is expected to be on the sub-percent level. 

Of course this theory now has one too many quark flavours, hence we must take the square-root of the fermion determinant that represents the $s/s'$ contribution to the Feynman path integral, which in the lattice context provides the weight of each gauge configuration in the ensemble average. This is a strategy commonly used for the staggered formulation of lattice QCD. Unfortunately taking the square-root results in a non-local determinant, although unlike for the staggered formulation the non-locality is confined to the boundaries and should be benign at sufficiently large volumes.

The use of GPBC impacts the forms of the diagrams involved in a given calculation: The mixing of quark flavours allows for the Wick contraction of up and down quark field operators resulting in non-zero values for the Green's functions:
\vspace{-0.3cm}
\begin{equation}
\prop^{(2,1)}_{y,x} = \langle C\bar u^T_y \bar d_x\rangle\,,\hspace{1cm}\prop^{(1,2)}_{y,x} = \langle -d_y u^T_x C^T \rangle\,.
\vspace{-0.2cm}
\end{equation}
This results in an increased number of diagrams that must be evaluated. In the first of these contractions, quark flavour flows towards the boundary on both sides. Likewise, quark flavour flows away from the boundary in the second contraction. We may interpret this as the boundary destroying/creating flavour, violating baryon number conservation. In practise this means that baryon-number eigenstates are not eigenstates of the system; for example the proton ($uud$) mixes with the anti-neutron ($\bar d\bar d \bar u$). However this is not important for calculations involving only mesonic states.

In order to demonstrate that the GPBC have the desired effect, we generated fully dynamical ensembles of domain wall fermions with a lattice volume of $16^3\times 32$ and an unphysically-large pion mass of $\sim 420$ MeV, with periodic BC and also GPBC in one and two directions. In figure~\ref{fig-dispersion} we plot the measured pion and kaon energies as a function of the number of directions with GPBC. We see clearly the increase in pion energy associated with the increasing number of G-parity boundaries, and that it agrees well with the expected dispersion relation. We also see that stationary kaon states can be produced in this framework. We also consider the quantity $B_K$, which measures the amplitude of mixing between neutral kaon states via the weak interaction. As it involves only kaons, we expect this quantity to be invariant under changing the number of G-parity boundaries; from the figure we see that this is indeed the case.

\vspace{-0.5cm}
\section{Conclusions and Outlook}
\vspace{-0.3cm}
After performing the calculation of the $\Delta I=1/2$ $K\rightarrow \pi\pi$ amplitude we will have all of the pieces required for a complete {\it ab initio} determination of the measure of direct CP-violation in the Standard Model. This calculation requires significant computational and theoretical advances to be made, particularly in the strategy used to obtain physical kinematics in the decay. Much of this work has now been completed, and the RBC and UKQCD collaboration are in the position to begin this measurement in earnest. To this end we have recently commenced the generation of a gauge ensemble with a lattice size of $32^3\times 64$, using M\"{o}bius domain wall fermions ($L_s=16$) and the Iwasaki+DSDR gauge action with $\beta=1.75$ and near physical pion masses (lightest unitary pion $171(1)$ MeV and lightest partially-quenched $143(1)$ MeV~\cite{Arthur:2012opa}). We are using G-parity boundary conditions in three directions to match the expected $K$ and $\pi\pi$ energies. (Note that aside from the 
boundary conditions, these lattice parameters are essentially the same as those used for the initial $\Delta I=3/2$ calculation.) The generation is being performed on 512 nodes of the IBM BlueGene/Q supercomputer at Brookhaven National Laboratory. We will soon have enough configurations to begin testing our measurement apparatus.

Our work now is focused upon formulating and testing the measurement strategy, including the technique we will use to evaluate the vacuum diagrams. Further testing of the systematic errors associated with the G-parity technique is also required, although thus far we have observed no evidence of any sicknesses. Further afield we might also consider the uses of G-parity boundary conditions in other frontier calculations performed by the collaboration, particularly those with significant noise contributions from intermediate pion states such as the calculation of the $K_L-K_S$ mass difference~\cite{Christ:2012se}.

\FloatBarrier

\end{document}